\documentclass[%
 reprint,
superscriptaddress,
 amsmath,amssymb,
 aps,prc
]{revtex4-2}

\usepackage{graphicx}% Include figure files
\usepackage{dcolumn}% Align table columns on decimal point
\usepackage{bm}% bold math
\usepackage[breaklinks=true]{hyperref}% add hypertext capabilities
\usepackage{color} %to highlight MMC changes
\usepackage[T1]{fontenc}

\usepackage{graphicx}
\usepackage{dcolumn}
\usepackage{bm}
\begin{document}

\title{High precision mass measurements of the isomeric and ground states of $^{44}$V: improving constraints on the IMME parameters of the $\text{A}=44$, $\text{0}^{\text{+}}$ quintet}
%Constraints the A=44 T=2 Isospin Quintet}


\author{D. Puentes}
\email[]{puentes@nscl.msu.edu}
\affiliation{Department of Physics and Astronomy, Michigan State University, East Lansing, Michigan 48824, USA}
\affiliation{National Superconducting Cyclotron Laboratory, East Lansing, Michigan 48824, USA}
\author{G. Bollen}
\affiliation{Department of Physics and Astronomy, Michigan State University, East Lansing, Michigan 48824, USA}
\affiliation{Facility for Rare Isotope Beams, East Lansing, Michigan 48824, USA}
\author{M. Brodeur}
\affiliation{Department of Physics, University of Notre Dame, Notre Dame, Indiana 49556, USA}
\author{M. Eibach}
\affiliation{Institut f\"ur Physik, Ernst-Moritz-Arndt-Universit\"at, 17487 Greifswald, Germany}
\author{K. Gulyuz}
\affiliation{Department of Physics, Central Michigan University, Mount Pleasant, Michigan 48859, USA}
\author{A. Hamaker}
\affiliation{Department of Physics and Astronomy, Michigan State University, East Lansing, Michigan 48824, USA}
\affiliation{National Superconducting Cyclotron Laboratory, East Lansing, Michigan 48824, USA}
\author{C. Izzo}
\altaffiliation[Current address: ]{TRIUMF, 4004 Wesbrook Mall, Vancouver, British Columbia V6T 2A3, Canada}
\affiliation{Department of Physics and Astronomy, Michigan State University, East Lansing, Michigan 48824, USA}
\affiliation{National Superconducting Cyclotron Laboratory, East Lansing, Michigan 48824, USA}
\author{S. M. Lenzi}
\affiliation{Dipartimento di Fisica e Astronomia ``Galileo Galilei'', Universit\'a di Padova, I-135131 Padova, Italy}
\affiliation{INFN, Sezione di Padova, I-35131 Padova Italy}
\author{M. MacCormick}
\affiliation{Universit\'e Paris-Saclay, CNRS/IN2P3, IJCLab, 91405 Orsay, France}
\author{M. Redshaw}
\affiliation{National Superconducting Cyclotron Laboratory, East Lansing, Michigan 48824, USA}
\affiliation{Department of Physics, Central Michigan University, Mount Pleasant, Michigan 48859, USA}
\author{R. Ringle}
\affiliation{National Superconducting Cyclotron Laboratory, East Lansing, Michigan 48824, USA}
\author{R. Sandler}
\affiliation{National Superconducting Cyclotron Laboratory, East Lansing, Michigan 48824, USA}
\affiliation{Department of Physics, Central Michigan University, Mount Pleasant, Michigan 48859, USA}
\author{S. Schwarz}
\affiliation{National Superconducting Cyclotron Laboratory, East Lansing, Michigan 48824, USA}
\author{P. Schury}
\affiliation{Wako Nuclear Science Center (WNSC), Institute of Particle and Nuclear Studies (IPNS), High Energy Accelerator Research Organization (KEK), Wako, Saitama 351-0198, Japan}
\author{N. A. Smirnova}
\affiliation{CENBG, CNRS/IN2P3 and Universit\'e de Bordeaux, Chemin du Solarium, 33175 Gradignan Cedex, France}
\author{J. Surbrook}
\affiliation{Department of Physics and Astronomy, Michigan State University, East Lansing, Michigan 48824, USA}
\affiliation{National Superconducting Cyclotron Laboratory, East Lansing, Michigan 48824, USA}
\author{A. A. Valverde}
\altaffiliation[Current address: ]{Physics Division, Argonne National Laboratory, Lemont, Illinois 60439, USA}
\affiliation{Department of Physics, University of Notre Dame, Notre Dame, Indiana 49556, USA}
\author{A. C. C. Villari}
\affiliation{Facility for Rare Isotope Beams, East Lansing, Michigan 48824, USA}
\author{I. T. Yandow}
\affiliation{Department of Physics and Astronomy, Michigan State University, East Lansing, Michigan 48824, USA}
\affiliation{National Superconducting Cyclotron Laboratory, East Lansing, Michigan 48824, USA}

%\thanks{}

\date{\today}

\begin{abstract}
\begin{description}
\item[Background]
The quadratic Isobaric Multiplet Mass Equation (IMME) has been very successful at predicting the masses of isobaric analogue states in the same multiplet, while its coefficients are known to follow specific trends as functions of mass number. The Atomic Mass Evaluation 2016 [Chin. Phys. C \textbf{41}, 030003 (2017)] $^{44}$V mass value results in an anomalous negative $c$ coefficient for the IMME quadratic term; a consequence of large uncertainty and an unresolved isomeric state. The $b$ and $c$ coefficients can provide useful constraints for construction of the isospin-nonconserving (INC) Hamiltonians for the $pf$ shell. In addition, the excitation energy of the $0^+, T=2$ level in $^{44}$V is currently unknown. This state can be used to constrain the mass of the more exotic $^{44}$Cr.   
\item[Purpose]
The aim of the experimental campaign was to perform high-precision mass measurements to resolve the difference between $^{44}$V isomeric and ground states, to test the IMME using the new ground state mass value and to provide necessary ingredients for the future identification of the $0^+$, $T=2$ state in $^{44}$V. 
\item[Method]
High-precision Penning trap mass spectrometry was performed at LEBIT, located at the National Superconducting Cyclotron Laboratory, to measure the cyclotron frequency ratios of [$^{44g,m}$VO]$^+$ versus [$^{32}$SCO]$^+$, a well-known reference mass, to extract both the isomeric and ground state masses of $^{44}$V.
\item[Results]
The mass excess of the ground and isomeric states in $^{44}$V were measured to be $-23\ 804.9(80)$ keV/$\text{c}^2$ and $-23\ 537.0(55)$ keV/$\text{c}^2$, respectively. This yielded a new proton separation energy of $S_p$ = 1\ 773(10) keV. 
\item[Conclusion]
The new values of the ground state and isomeric state masses of $^{44}$V have been used to deduce the IMME $b$ and $c$ coefficients of the lowest $2^+$ and $6^+$ triplets in $A=44$. The $2^+$ $c$ coefficient is now verified with the IMME trend for lowest multiplets and is in good agreement with the shell-model predictions using charge-dependent Hamiltonians. The mirror energy differences were determined between $^{44}$V and $^{44}$Sc, in line with isospin-symmetry for this multiplet. The new value of the proton separation energy determined, to an uncertainty of 10 keV, will be important for the determination of the $0^+$, $T=2$ state in $^{44}$V and, consequently, for prediction of the mass excess of $^{44}$Cr. 
\end{description}
\end{abstract}

\maketitle

%\tableofcontents

\section{Introduction}
Isospin is a quantum number that was postulated by Heisenberg after the discovery of the neutron \cite{Chadwick1932} to explain symmetries between the new nucleon and the proton \cite{Heisenberg1932}. In this formalism a nucleon is assumed to carry an isospin quantum number $t{=}1/2$, similar to an ordinary spin, with  proton and neutron being labeled by its projection $t_z{=\pm}1/2$. Therefore, the three components of the isospin operator, $\hat t$, obey well-known SU(2) commutation relations: $[\hat t_j,\hat t_k]=i \epsilon_{jkl} \hat t_l$. The total isospin operator for an $A$-nucleon system is $\hat T=\sum_{n=1}^A \hat t(n)$, while its projection is $\hat T_z=\sum_{n=1}^A \hat t_z(n)$. A charge-independent Hamiltonian would commute with isospin operator, $[\hat H,\hat T]=0$, giving rise to degenerate multiplets of states $(J^{\pi },T)$ in nuclei with the same $A$ and $T_z=-T,\ldots,T$, called {\it isobaric analogue states (IAS)}.

However, the Coulomb interaction between protons, the proton-neutron mass difference, and small charge-dependent components in phenomenological nuclear interactions lead to an energy splitting of the isobaric multiplet states. Wigner showed \cite{Wigner1957}  that, assuming a two-body nature of charge-dependent forces, the splitting of the isobaric multiplets follows the quadratic dependence as a function of $T_z$, 
%isospin could be used to predict the binding energies of nuclei with great success for light nuclei, 
establishing what is known as the Isobaric Multiplet Mass Equation (IMME):
\begin{equation} \label{IMME}
M.E.(\alpha ,T) = a(\alpha ,T) + b(\alpha ,T) T_z+c(\alpha ,T)T_z^2 \,,
\end{equation}
where $\alpha $ denotes all other relevant quantum numbers which characterize a given multiplet ($A$, $J$, $\pi $, $\ldots$).

The quadratic IMME describes well the energy splittings of the isobaric multiplet states \cite{Wigner1957}. For quartets and quintets, it is often used to determine masses of unknown members belonging to the same multiplet. Data from high-precision mass measurements and other experimental techniques have made it possible to work out extensive compilations of the experimental $a$, $b$ and $c$ coefficients for the lowest and higher lying multiplets up to about $A=71$ (the most recent evaluations can be found in Refs.~\cite{Lam2013, MacCormick2014}).  Some intriguing results are unveiled from the behavior of the IMME coefficients as a function of $A$, which manifests in specific trends and even characteristic fine structure (staggering) for $T=1/2,1,3/2$.
In addition, any manifestation of a deviation from a quadratic IMME in quartets or quintets is also of high interest since it could either be related to isospin mixing or bring important information on the presence of charge-dependent many-body forces~\cite{Janecke1969,Bertsch1970,Benenson1979}.

Understanding the $a$, $b$ and $c$ coefficients is important for nuclear structure theory, since they serve as a probe of the strengths of the isoscalar, isovector and isotensor components of the nuclear force, respectively, and therefore can shed light onto the magnitude of the charge-dependent terms of the effective nucleon-nucleon interaction in different approaches~\cite{Janecke1966,Ormand1989,Nakamura1994,Bentley2007,kaneko2012,kaneko2013,LamSm2013,KlSm2020}. While the main contribution comes from the Coulomb interaction between the charged protons, the charge-dependent components of nuclear origin are not negligible and have to be taken into account for a realistic description. 

A precise description of the IMME coefficients remains challenging for nuclear theory and numerous studies are underway. For lighter, $sd$ and $pf$ shell, nuclei the most accurate predictions are due to the shell-model with phenomenological isospin-nonconserving (INC) Hamiltonians~\cite{Ormand1989,LamSm2013}. To construct such a Hamiltonian, one adds to a well-established isospin-conserving Hamiltonian, such as USD \cite{Brown1988,Brown2006} in the $sd$ shell, a charge-dependent part, consisting of the two-body Coulomb interaction, an isovector and an isotensor term describing effective charge-dependent two-body forces of nuclear origin, and isovector single-particle energies. The strengths of those terms are adjusted to match the experimentally extracted IMME $b$ and $c$ coefficients. For the $sd$ shell this task was accomplished in the 1990's \cite{Ormand1989,Nakamura1994} and recently updated \cite{LamSm2013}, using a more extensive database. There has been less progress with the $pf$ shell, however. In Ref.~\cite{Ormand1989}, while the isovector strength parameter in the $pf$ shell was obtained by a fit of the $b$ coefficients, the strength of the isovector component of nuclear origin was estimated to be approximately 4\% of the $T=1$ two-body matrix elements. 
One reason for that is the difficulty of the calculations at that time, while the other reason was the lack of experimental data.

Modern progress in the shell-model computations gave rise to new higher precision effective interactions for large-scale calculations in the full $pf$ shell, such as GXPF1A~\cite{Honma2004} or KB3G~\cite{Poves2001}. 
The INC version of GXPF1A, which we call here cdGX1A, still adapts the charge-dependent terms parametrization from Refs.~\cite{Ormand1989,Ormand1995}. The recently derived microscopic INC  $pf$ shell Hamiltonians~\cite{Ormand2017}, based on the modern $NN$ potentials and many-body perturbation theory, show themselves to be less successful than earlier phenomenological parametrizations. It would be interesting to further test the predictions of phenomenological INC interactions on an updated experimental database, and, if necessary, to work out a well-adjusted $pf$ shell INC interaction.

%There have been attempts at using the INC Hamiltonian in large scale $0\hbar \omega$ calculations, and modern versions of the KB3G and GXPF1A calculations have been performed \cite{Lam2013}. The studies were met with limited success, however, due to different factors, such as a lack of experimental data to use for inputs in the calculations. Another issue is the large isotensor Coulomb strength parameters that comes from fitting ground and exited states in the $fp$ shell \citep{Lam2013}. There is also the possibility that the charge-dependent form of the Hamiltonian is incomplete.

A high-precision INC Hamiltonian would provide a framework to understand the structure and decay modes of proton rich nuclei, including mass predictions for nuclei in the vicinity of the proton drip-line. This would have consequences for nuclear astrophysics (e.g., for the $rp$-process). Another important application of such interactions is the calculation of nuclear structure corrections to superallowed Fermi $\beta $-decay between isobaric analogue states, used for the tests of the CVC hypothesis and the CKM matrix unitarity tests~\cite{Ormand1995,Hardy2015,SmirnovaNTSE2018}. 

\begin{figure*}
\includegraphics[scale=.5]{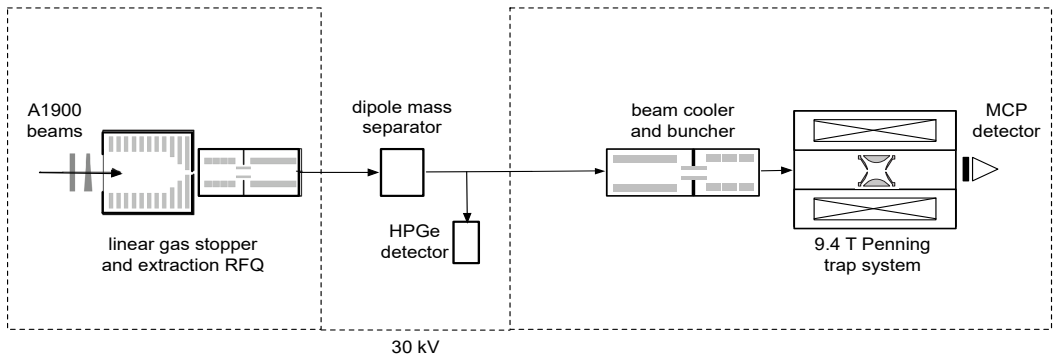}
\caption{Schematic diagram of the experimental setup including the beam stopping area, low energy transport/mass separation, LEBIT cooler/buncher and 9.4 T Penning Trap system.}
\label{fig:lebit}
\end{figure*} 

A full analysis of IMME coefficients for the ground state of nuclei completed for $T = 1/2$ to $3$ multiplets was performed~\cite{MacCormick2014} using the Atomic Mass Evaluation (AME) 2012 and the Evaluated Nuclear Structure Data Files \cite{Audi2012a,Audi2012b}. There were different cases where unknown mass values were the largest source of uncertainty. For the $T=1$ triplet at $A = 44$, the measured ground state of $^{44}$V  produced an improbable negative $c$ coefficient in the IMME. At the same time, the predicted value from the $c$ coefficient systematics was around 145~keV~\cite{MacCormick2014}. If it were true, it would point to the possibility of new physics. With such a large uncertainty, new high-precision measurements are necessary to resolve the discrepancy in the $c$ coefficient and determine how well the INC-Hamiltonian describes the $A=44$ case. This creates an anchor for the $A = 4n$ nuclei (where $n$ is an integer) and provides the information necessary to formulate a complete $sdpf$ shell model description of light nuclei.

In the last two decades, significant progress has been made in understanding the Coulomb energy shifts in isobaric analogue states for both mirror energy differences (MEDs) and triplet energy differences (TEDs). These quantities for doublets and triplets are related to the differences in the $b$ and $c$ coefficients of the excited and lowest multiplets: MED$_J=-2\Delta b_J$,  TED$_J=2\Delta c_J$ (the latter is not defined for  $T=1/2$ doublets). The theoretical description of MEDs and TEDs in the $pf$ shell using an INC Hamiltonian for the $pf$ shell has been a great success~\cite{Bentley2007}. A particular feature of the approach is the use of an effective two-body component, of nuclear origin, parametrized in the form of a few specific two-body matrix elements. For the $pf$ shell, there was evidence that an additional isovector two-body matrix element for the $f_{7/2}$ sub-shell in the MED was needed. In the TED, there is evidence that an additional isotensor two-body matrix element is necessary for the $f_{7/2}$ sub-shell. There was also a study of $^{44}$V showing that the $J=2$ matrix elements differ from the other $T=1$ multiplets as described by theory~\cite{Taylor2011}. Application of that description to the $A=44$ case was difficult since the position of the $6^+$ isomer was not known. The high-precision measurement of the $6^+$ $^{44}$V isomeric state provides experimental constraints on both the MED and TED studies in this region~\cite{Taylor2011}.  

Due to a lack of experimental data for the $^{44}$Cr nucleus, the $0^+$ quintet group in $A = 44$ is even less well known~\cite{Wang2017}. Precise knowledge of the $^{44}$V mass is required for an accurate determination of the mass excess of the $0^+$, $T = 2$ state, an IAS of the $^{44}$Cr ground state. This will provide the first experimental constraint on the ground-state binding energy of $^{44}$Cr. The $\beta $-delayed proton decay of $^{44}$Cr has been observed experimentally~\cite{Dossat2007,Pomorski2014}, but the assignment of the proton branch from the IAS was tentative. A more precise value of the one-proton separation energy in $^{44}$V will contribute to future studies aimed at identifying the exact location of the proton-emitting level, to establish whether it is from the IAS or not, and will also be useful in pinning down the positions of other proton-unbound levels.

Recently, a mass measurement of $^{44}$V was performed using isochronous mass spectrometry at the Cooler Storage Ring (CSRe), Lanzhou~\cite{Zhang2018}. They were able to resolve the ground and isomeric state of $^{44}$V, and found a new $c$ coefficient that agreed with the predicted IMME parameters. Through a high-precision measurement using Penning trap mass spectrometry, we are in a position to verify their mass measurements, which currently disagrees by 1.7$\sigma$ with the AME2016~\cite{Wang2017} for the ground state and 1.7$\sigma$ for the isomeric state using the prediction from \cite{Taylor2011} with the AME2016 value.
%It was unknown which level was the T=2 due to the large uncertainty of the ground state of $^{44}$V and the density of levels near that excitation energy.  
\section{Experimental Methods}
At the National Superconducting Cyclotron Laboratory (NSCL), a radioactive $^{44}$V beam was produced through projectile fragmentation. A stable $^{58}$Ni beam was accelerated using the Coupled Cyclotron Facility (CCF) to an energy of 160 MeV/nucleon, and impinged on a $^{9}$Be target with a thickness of 705 mg/cm$^2$. After the target station, the cocktail beam was purified in the A1900 Fragment Separator~\cite{Morrissey2003}.

The beam energy was degraded in a 1530 $\mu$m thick Al plate and a 1050 $\mu$m thick glass silica wedge to $\approx$ 1 MeV/nucleon~\cite{Morrissey2003}. The beam was then delivered to a gas cell, filled with high purity He buffer gas at a pressure of 71 mbar, to stop the radioactive beam~\cite{Sumithrarachchi2020}. The ions were extracted through a Radio-Frequency Quadrupole (RFQ), accelerated to an energy of 30 keV, and purified using a magnetic dipole mass separator based on the mass to charge ratio ($A/Q$). The resolving power of the mass separator is $\sim$1500. The thermalized beam showed the most activity at $A/Q = 60$, corresponding to [$^{44}$VO]$^+$ molecular ion.

\begin{table*}[t]
\caption{\label{table} Mean frequency ratios, $\bar{R}=\nu_{c}$/$\nu_{c,\text{ref}}$, calculated atomic mass and mass excess (ME) values, and their comparison to the values from Ref~\cite{Zhang2018}. The uncertainty reported in curly brackets for the $\bar{R}$ are the inflated uncertainties after taking into account the Birge Ratio.}
\begin{ruledtabular}
\begin{tabular}{ccccccc}
% Lines of table here ending with \\
\multicolumn{1}{c}{Isotope} & \multicolumn{1}{c}{Measured Ion} & \multicolumn{1}{c}{Reference} & \multicolumn{1}{c}{$\bar{R}$} & \multicolumn{1}{c}{Mass ($^{44}$V) (u)} & \multicolumn{1}{c}{ME ($^{44}$V) (keV)}  & \multicolumn{1}{c}{Ref.~\cite{Zhang2018} (keV)} \\
\hline
$^{44}$V & $^{44}$VO$^{+}$ & [$^{32}$SCO]$^+$ & $0.999\ 960\ 43(13)\{14\}$ & $43.974\ 444\ 1(84)$ & $-23\ 804.9(80)$  &  $-23\ 827(20)$ \\
$^{44m}$V & $^{44m}$VO$^{+}$ & [$^{32}$SCO]$^+$ & $0.999\ 955\ 631(75)\{98\}$ & $43.974\ 731\ 9(57)$ & $-23\ 537.0(55)$  & $-23\ 541(19)$\\
\end{tabular}
\end{ruledtabular}
\end{table*}

\begin{figure}
\centering
\includegraphics[scale=0.35]{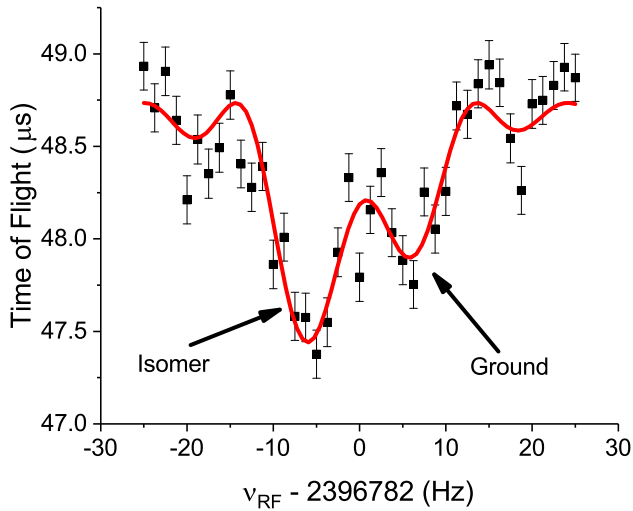}
\caption{A 100-ms [$^{44}$VO]$^+$ time-of-flight cyclotron double resonance used for determining the $\nu_c$ of [$^{44}$VO]$^+$. The line shape is a theoretical curve described in \cite{Konig1995}. Both the isomer and the ground state of $^{44}$V are visible.}
\label{fig:Resonance}
\end{figure}

The low-energy beam was then guided through an electrostatic beam transport system to the Low Energy Beam and Ion Trap (LEBIT) experimental station~\cite{Ringle2013}, as shown in Fig. \ref{fig:lebit}. The first component of the LEBIT station is a two-stage cooler and buncher \cite{Schwartz2016}. The cooler stage uses He buffer gas to continue thermalizing the beam while the buncher converts the DC beam into a pulsed beam. The ions are then ejected into the 9.4 T Penning Trap. Once trapped, the ions were first purified using narrow-band and SWIFT rf dipole excitations~\cite{Blaum2004,Kwiatkowski2015} via segmented ring electrodes on the trap to remove contaminant ions. The Time-of-Flight Ion Cyclotron Resonance (TOF-ICR) technique was used to measure the cyclotron frequency, $\nu_c= qB/(2\pi\cdot m)$, of an ion of interest of mass $m$ and charge $q$ in a magnetic field $B$ ~\cite{Bollen1990,Konig1995}. A TOF response curve is shown in Fig. \ref{fig:Resonance}, where both the isomer and ground state are resolved. The data were fitted with a theoretical line shape from which the cyclotron frequency was extracted~\cite{Konig1995}. All measurements of [$^{44}$VO]$^+$ were performed using a 100 ms excitation time, whereas the reference ion ([$^{32}$SCO]$^+$) was measured using a 500 ms excitation time.

\section{Results} 
The frequency ratio of the ion of interest $\nu_{c}$ to the reference ion $\nu_{c,\text{ref}}$, $\bar{R}=\nu_{c}$/$\nu_{c,\text{ref}}$, is used to determine the mass of the ion of interest. This ratio is obtained by alternating measurements of the reference ion and the ion of interest. Whenever the cyclotron frequency of [$^{44}$VO]$^+$ was measured, there was also an associated measurement of the cyclotron frequency of the reference ion, [$^{32}$SCO]$^+$, taken before and after. The reference ion was measured with a longer excitation time to ensure that the uncertainty in the frequency of the reference ion does not contribute significantly to the uncertainty of the ratio. These two reference measurements are used to interpolate the value of $B$ at the time [$^{44}$VO]$^+$ was measured.  In this manner a ratio of the frequency of interest to the frequency of the well-known reference ion is determined.  After repeating this cycle multiple times the mean frequency ratio $\bar{R}$ is obtained.  

Measurements were taken over an 80-hour period. Due to low statistics, multiple measurements were compiled into five combined resonance curves.  The mean frequency ratio for the ground state of [$^{44}$VO]$^+$ vs [$^{32}$SCO]$^+$ was measured to be $\bar{R}$ = 0.999\ 960\ 43(13) with a Birge ratio \cite{Birge1932} of 1.07(21). The mean frequency ratio for the isomeric state was measured to be $\bar{R}$ = 0.999\ 955\ 631(75) with a Birge ratio of 1.31(21). Since the Birge ratios were greater than 1, the uncertainties of $\bar{R}$ were inflated to account for potential systematic uncertainties.

The mass excess of $^{44}$V was extracted using 
\begin{equation}
m_{V}=\cfrac{1}{\bar{R}}(m_{S}-m_e )+m_e
\end{equation}
where $m_{V}$ is the mass of [$^{44}$VO], $m_{S}$ is the mass of [$^{32}$SCO] and $m_e$ is the mass of the electron. $m_{S}$ is determined from the AME2016 mass values of its constituent isotopes~\cite{Wang2017}. The molecular binding energy and ionization energy of the singly-charged molecular ion are on the order of eV or less and are not included.

The measured effect from previous work of nonlinear magnetic field fluctuations was on the order of $3.0 \cdot 10^{-10}/\text{u}/\text{hour}$ \cite{Ringle2007b}. The longest measurement, after combining measurements due to low statistics, was 12 hours. However, since the reference ion and ion of interest are similar in mass, the effect of magnetic field fluctuations is negligible. A large number of ions in the trap can lead to frequency shifts due to the Coulomb interaction between the trapped ions. When measurements are performed with many ions in the trap (e.g. five or more), a count (z) class analysis~\cite{kellerbauer2003} is used to extrapolate the frequency of a single ion in the trap. Data is analyzed based on the number of ions in the trap (a count class), followed by a fit of each class to determine a frequency. The frequency of a single ion in the trap is then extrapolated from a linear regression of this data. A count class analysis was not necessary due to the low statistics of the experiment because events with only five ions or fewer were used for the analysis. Finally, due to the similarity in mass between the reference ion and ion of interest, relativistic effects did not need to be considered.

The results of the data analysis are shown in Table \ref{table}. These values agree with the values reported in \cite{Zhang2018}, but are a factor of 2-4 more precise. Our new mass values also disagree with the AME2016 by 1.6$\sigma$ for the ground state and by 1.7$\sigma$ for the isomeric state. The AME2016 value was adopted from the AME2012 value, which was evaluated from \cite{Stadlmann2004}. The $^{44}$V mass excess was determined by using experimental fit parameters obtained during the experimental ring storage experiment reported by \cite{Stadlmann2004}. A new proton separation energy for $^{44}$V is determined to be 1\ 773(10) keV. 

\section{Discussion}

Previous work from \cite{Taylor2011} established a level scheme based on measured gamma-rays. However, as the energy of the isomeric state was unknown, many excited state energies of $^{44}$V were determined relative to the isomeric state. Using our new measurement reported here, the excitation energies were experimentally determined with uncertainties based on the energy uncertainty of the isomeric state. Values for the  $^{44}$V excited state energies are listed in Table \ref{tab:ExcitedStates} compared to predictions from shell model calculations. These values were used to calculate the experimental MEDs and TEDs, which are explained in further detail in subsection \ref{MEDTED}.

\begin{table}[htb]
\caption{Newly constrained levels of $^{44}$V based on the resolved isomeric state. This work shows the experimental data while \cite{Taylor2011} is based on shell-model calculations, with the only uncertainty coming from the gamma rays observed in Ref. \cite{Taylor2011}}
\begin{ruledtabular}
\begin{tabular}{ccc}
 $J^{\pi}$ & This work + Ref. \cite{Taylor2011} (keV) & Ref. \cite{Taylor2011} (keV) \\
\hline
(7$^+$) & 981.7(98) & 979.9(5) \\
(9$^+$) & 2\ 661.7(99) & 2\ 659.9(17) \\ 
(10$^+$) & 4\ 027(10) & 4\ 024.9(19) \\
(11$^+$) & 3\ 494(10) & 3\ 491.9(18) 
\end{tabular}
\end{ruledtabular}
\label{tab:ExcitedStates}
\end{table} 

\subsection{IMME b and c coefficients of the lowest $A=44, T=1$ multiplets}

\begin{figure}
\centering
\includegraphics[scale=0.465]{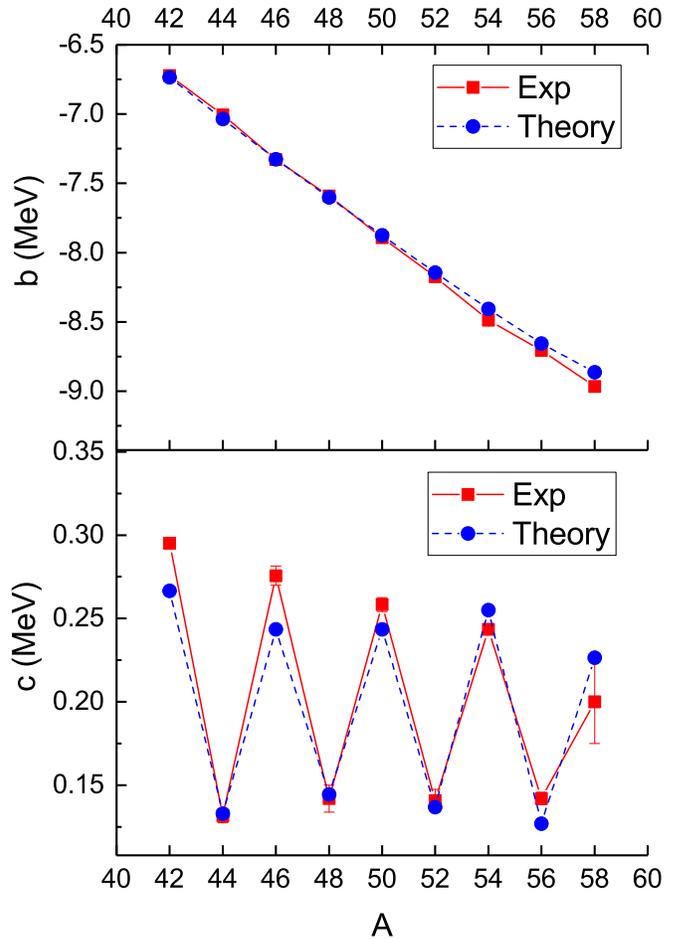}
\caption{Experimental (square) and theoretical (circle) $b$ and $c$ coefficients of the lowest triplets in the $pf$ shell, as a function of mass number $A$. The experimental information came from references \cite{Wang2017,Dossat2007,Pomorski2014,Valverde2018} and this work on $A=44$ ground state mass measurements. Theoretical calculations have been performed with cdGX1A Hamiltonian~\protect\cite{NuShellX}.}
\label{fig:coeff}
\end{figure}

The new values of the mass excesses of $^{44}$V and $^{44m}$V complete the set of experimental data for the lowest $2^+$ and $6^+$, $T=1$ multiplets in $A=44$ and enable the deduction of new values of their respective IMME $b$ and $c$ coefficients using Eq. \ref{IMME}. The $b$ coefficient for the $2^+$, $T=1$ triplet is deduced to be $b=-7005.5(41)$~keV, while the new value for its $c$ coefficient is $131.75(420)$ keV. As seen from Fig.~\ref{fig:coeff}, which depicts IMME $b$ and $c$ coefficients for the lowest $pf$ shell triplets, both values for $A=44$ match well the known systematics in this mass region. This resolves the anomaly previously existing at $A=44$ and verifies the work performed by~\cite{Zhang2018}. The experimental $b$ and $c$ coefficients for the lowest $6^+$ triplet have values $-7003.9(2.7)$ keV and $158.9(2.8)$ keV, respectively.

\begin{table}[htb]
\caption{New experimental and theoretical IMME $b$ and $c$ coefficients for the lowest $2^+$ and $6^+$ triplets in $A=44$.}
\begin{ruledtabular}
\begin{tabular}{ccccc}
$A=44$ & \multicolumn{2}{c}{$2^+$} & \multicolumn{2}{c}{$6^+$} \\
\hline
 & $b$ & $c$ & $b$ & $c$ \\
& keV & keV & keV & keV \\
\hline
Exp & -7\ 005.5(41)  & 132(5) & -7\ 003.9(27) & 159(3) \\
cdGX1A & -7\ 035 & 133 & -7\ 025 & 153 \\ 
cdFPD6 & -7\ 037 & 132 & -7\ 019 & 146
\end{tabular}
\end{ruledtabular}
\label{tab:coefficients}
\end{table} 

To understand these values and to test the existing theoretical models, we performed shell-model calculations in the full $pf$ shell using NuShellX@MSU~\cite{NuShellX} code with two different charge-dependent Hamiltonians, cdGX1A and cdFPD6. These Hamiltonians are based on the isospin-conserving Hamiltonians, GXPF1A~\cite{Honma2004} and FPD6~\cite{FPD6}, respectively, to which have been added the two-body Coulomb interaction, effective nuclear charge-symmetry breaking and charge-independence breaking terms from Ref.~\cite{Ormand1989}, and isovector single-particle energies from Ref.~\cite{Ormand1995}. These interactions can be found in the NuShellX@MSU package. A scaling factor proportional to $\sqrt{\hbar \omega (A)}$ is imposed on the Coulomb interaction and isovector single-particle energies to account for the mass-dependence throughout the shell.

Theoretical values of the IMME $b$ and $c$ coefficients for the lowest $2^+$ and $6^+$ multiplets in $A=44$, as obtained with cdGX1A and cdFPD6, are given in Table \ref{tab:coefficients}. Regarding $c$ coefficients, there is a remarkable agreement of theory with the data, especially for the $2^+$ multiplet. For $b$ coefficients, we observe more discrepancy between theory and experiment, of the order of 20-30 keV. This is roughly the average precision currently reached for the $pf$ shell. To demonstrate, we show in Fig.~\ref{fig:coeff} the cdGX1A predictions for the IMME $b$ and $c$ in the $pf$ shell from $A=42$ to $A=58$. The root-mean-square deviation for $b$ coefficients is 48 keV, for $c$ coefficients is 17 keV.
We conclude that there is very good agreement of the current phenomenological description with experiment.
%This provides excellent agreement with isospin symmetry conservation. 
%Despite not being a part of the isospin multiplet, a similar analysis was performed for the Isomeric state.

The isomeric $6^+$ state in $^{44}$V is found by cdGX1A to be the first excited state 
at 282~keV (cdGX1A) above the $2^+$ ground state.
cdFPD6 is less predictive: the $6^+$ state is found to be the second cdGX1A excited state at
643~keV (cdFPD6) above the $2^+$ ground state, with the $4^+$ state being slightly below.
As seen from Table \ref{tab:coefficients}, the calculated values of the $b$ and $c$ coefficients using either Hamiltonian are in fair agreement with the experimental value. In spite of these already encouraging results, it will definitely be beneficial in future construction of the INC interactions of improved precision to take into account the $A=44$ data on $b$ and $c$ coefficients.
%The disagreement in predicted levels compared to experimental data, stress the importance of the parametrization of the charge-dependent terms.

\subsection{Mirror and triplet energy difference} \label{MEDTED}
\begin{table}[b]
   \caption{A comparison of the experimental MED for the $T=1$, $J=6$ isomeric state in $^{44}$V-$^{44}$Sc mirror nuclei with the theoretical value obtained  using Eq.~\eqref{MED}.}
    \begin{ruledtabular}
    \begin{tabular}{cc}
    % Lines of table here ending with \\
    \multicolumn{1}{c}{Parameter} & \multicolumn{1}{c}{Value} \\
    \hline
    Experimental MED (keV) & -3(10) \\
    Theoretical MED (keV) & -3.3 \\
    $V_{CM}$ (keV) & -26 \\
    $V_B$ (keV) & -1.3 \\ 
    $V_{Cm}$ (keV) & +24 \\
    \end{tabular}
    \end{ruledtabular}
    \label{tab:MED}
\end{table}

Differences between the excitation energy of analogue states in mirror nuclei, MEDs, can now be deduced and compared with the shell-model calculations using INC interactions.
\begin{equation}
     {\rm MED(J)} = E^*_{T_z=-1}(J)- E^*_{T_z=+1}(J) \end{equation}{}

Following Refs.~\cite{Lenzi2001, Bentley2007}, we consider the contribution of three charge-dependent terms: the multipole Coulomb term that results from the expectation value of the Coulomb potential in both nuclei at each value of $J$ ($V_{CM} (J)$), and other two terms of monopole origin ($V_{Cm}(J)$). The first one is related to the change of the nuclear radius as a function of the spin and the second is related to corrections to the single-particle energies of protons and neutrons~\cite{Bentley2007}.  Systematic studies of nuclei in the $f_{7/2}$ shell have shown the need of adding an isospin-symmetry breaking term ($V_B(J)$) originally deduced from the experimental MED values ($^{42}$Ti-$^{42}$Ca) for two protons and two neutrons in the $f_{7/2}$ shell~\cite{Zuker2002}, and later extended to all active orbits~\cite{Bentley2015,Boso2018}. It consists of an isovector correction to the matrix-element for two nucleons coupled to $J=0, T=1$ with a strength of 100 keV smaller for protons than for neutrons. In this work we have taken into account the four orbits of the $pf$ shell. 
Theoretical calculations were performed using the GXPF1A effective interactions in the full $pf$ shell.

\begin{figure}
\centering
\includegraphics[scale=.48]{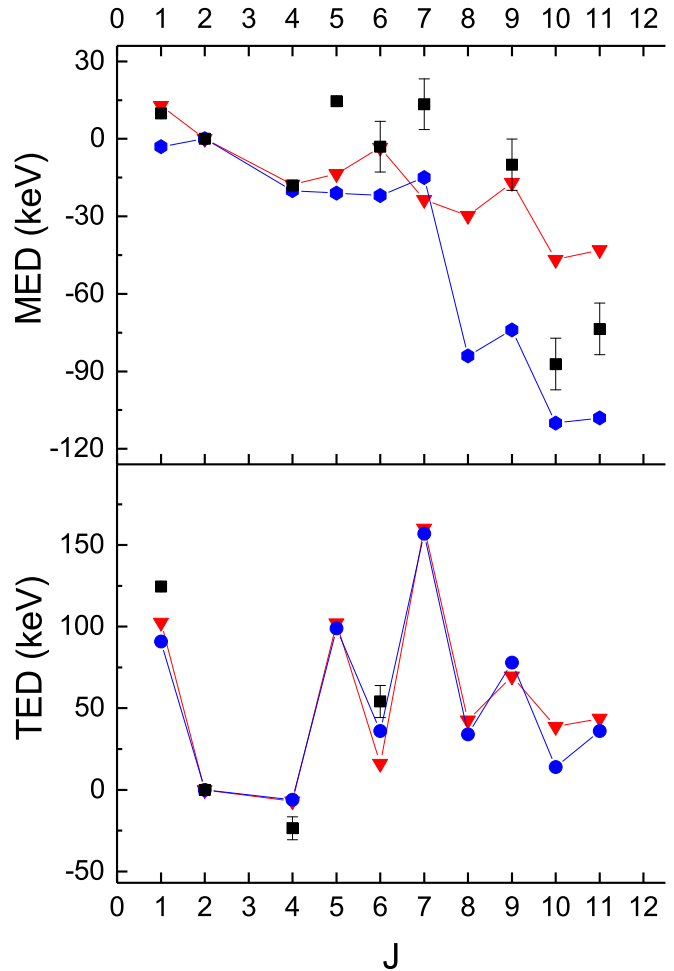}
\caption{Mirror and triplet energy differences as a function of J for $A = 44$. Experimental values (square) are compared with those obtained with two different shell model calculations: (circle) with the cdGX1A effective interaction ~\protect\cite{NuShellX} and (triangle) using Eqs.~(4,6) ~\cite{Bentley2007}. See text for details.}
\label{fig:MED-TED}
\end{figure}

The MED are calculated through the summation of all the potential contributions ($\Delta_M$ stands for the difference between the mirror states):
 \begin{equation}\label{MED}
     {\rm MED(J)} = \Delta_M V_{CM}(J) + \Delta_M V_{Cm}(J) + \Delta_M V_B(J)
 \end{equation}{}
Table~\ref{tab:MED} summarizes the experimental MED for the $6^+$ state in comparison to the MED obtained using Eq.~(\ref{MED}). The results show that the contribution from the $V_B$ term is negligible and, interestingly, the small MED results from the mutual cancellation of the monopole ($V_{Cm}$)  and multipole Coulomb ($V_{CM}$) terms. 
The increase of the monopole contribution with spin is interpreted as a shrinking of the radius with increasing angular momentum. This can be deduced from the shell model wave functions looking at the dramatic decrease of the occupation of the $p$ orbits, that have larger radius than the $f$ orbits in the $pf$ shell, as explained in~\cite{Lenzi2001,Bentley2007}. On the other hand, the decrease of the multipole contribution is due to the reduction of Coulomb repulsion through the alignment of protons in $^{44}$V at $J=6$. These two effects are of the same magnitude but of opposite sign, and cancel out at the isomeric state.

This analysis can be extended to other yrast excited states reported in Ref.~\cite{Taylor2011}. Yrast states are the lowest-energy states for given angular momentum~\cite{Robb1967}. Together with the  triplet energy differences, TEDs, among the $T=1$ isobaric analog states in $^{44}$V, $^{44}$Sc and $^{44}$Ti:
\begin{equation}\label{TEDex}
     {\rm TED(J)} = E^*_{T_z=-1}(J)+ E^*_{T_z=+1}(J) - 2 E^*_{T_z=0}(J) 
\end{equation}{}
Theoretically, these are obtained following \cite{Zuker2002,Bentley2007}:
 \begin{equation}\label{TED}
     {\rm TED(J)} = \Delta_T V_{CM}(J) + \Delta_T V_B(J),
 \end{equation}{}\noindent
where $\Delta_T$ indicates the difference between the triplet states as in eq.~(\ref{TEDex}). We note that in the TED the monopole contributions cancel out by construction. The $V_B$ term consists of an isotensor correction to the matrix-element for two nucleons coupled to $J=0, T=1$ with a strength of 100 keV.
MED and TED experimental values for the yrast states in $A=44$ are shown in Fig.~\ref{fig:MED-TED}. They are compared to the calculations using Eq.~\eqref{MED} and Eq.~\eqref{TED}, respectively, and with the cdGX1A interaction.
Both theoretical results are in fairly good agreement with data. For the MED, the difference between both theoretical approaches reside mainly in the monopole correction. These differences disappear in the TED where both methods give similar predictions.
 
\begin{figure}
\centering
\includegraphics[scale=0.6]{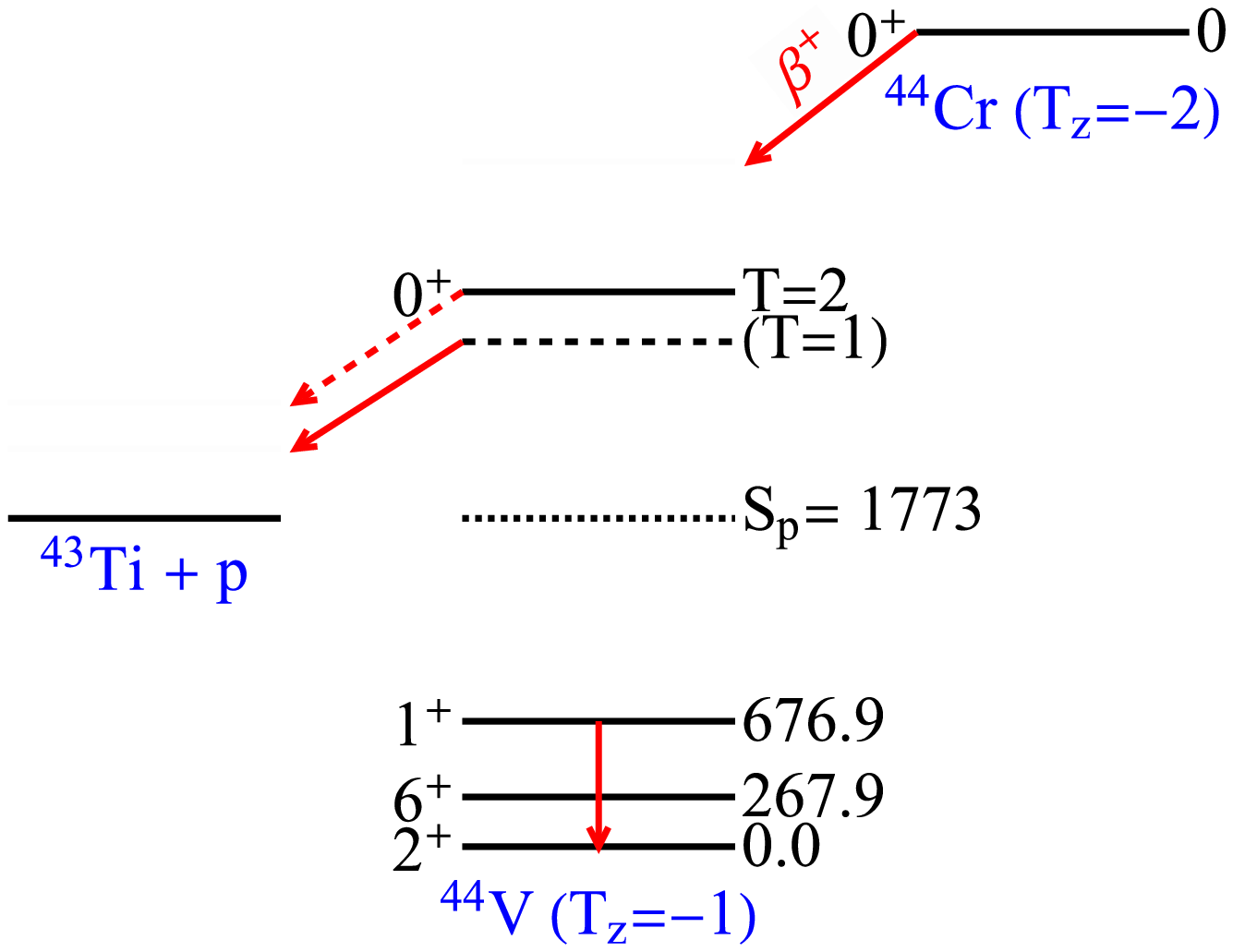}
\caption{Experimental partial decay scheme of $^{44}$Cr, including new results calculated with the new precise mass measurements of the ground and isomeric states of $^{44}$V. The experimental information came from references \cite{Wang2017,Dossat2007,Pomorski2014}. The level denoted $(T=1)$ refers to the unknown $J=1^+, T=1$ state (or states) that can undergo proton emission, in the vicinity of the IAS. The red dashed transition represents the isospin-forbidden transition to $^{43}$Ti.}
\label{fig:Levels}
\end{figure}

\subsection{Constraints for the ground state mass of $^{44}$Cr}

A new mass excess value of the ground state of $^{44}$V provides a definitive value for a one-proton separation threshold in this nucleus: $S_p=1773(10)$~keV. This new value is an important ingredient for the determination of the excited states based on the energies of protons emitted from $^{44}$V.
Indeed, $\beta$ decay of $^{44}$Cr populates proton-unbound states of $^{44}$V, one of which is the $0^+,T=2$ IAS of the $^{44}$Cr ground state, as shown in Fig. \ref{fig:Levels}.
$\beta $-delayed proton emission from $^{44}$Cr has been observed experimentally~\citep{Dossat2007,Pomorski2014}. No firm assignment of the $0^+,T=2$ state exists. Some of the observed peaks could belong to the proton emission from the Gamow-Teller populated $1^+,T=1$ states.
Future experiments on $\beta $-delayed spectroscopy of $^{44}$Cr should be able to at least pin down the excited level of the $0^+,T=2$ IAS in $^{44}$V, and using the new mass value, constrain the mass of $^{44}$Cr with greater precision. 

\section{Conclusions}

The first Penning trap mass measurements of $^{44}$V ground and isomeric states were performed. The two states were resolved in the Penning trap, lowering the mass uncertainties of these two states from the values reported previously~\cite{Zhang2018} by a factor of 2-4. This provided a lens to probe the improbable value of the IMME $c$ coefficient for the $A=44$, $T=1$ multiplet. The new values of the $b$ and $c$ coefficients extracted for the $2^+$ and $6^+$ triplets are in good agreement with the known systematics and with the IMME. The experimental MEDs and TEDs also support isospin symmetry the $A=44$, $T=1$ triplet. A new value for the proton separation threshold in $^{44}$V has been determined to an uncertainty of 10 keV. This should help identify the IAS in future experiments on $\beta$-delayed proton spectroscopy of $^{44}$Cr. Finally, the mass measurement of $^{44}$Cr would provide complete information on the isobaric quintet ($T=2$) and check the possibility of expanding the IMME to higher terms. 

\section{Acknowledgements}

This work was conducted with the support of Michigan State University, the National Science Foundation under Contracts No. PHY-1565546, No. PHY-1713857, No. PHY-1913554, and the U.S. Department of Energy, Office of Science, Office of Nuclear Physics under Award No. DE-SC0015927. The work leading to this publication has also been supported by a DAAD P.R.I.M.E. fellowship with funding from the German Federal Ministry of Education and Research and the People Programme (Marie Curie Actions) of the European Union's Seventh Framework Programme FP7/2007/2013 under REA Grant Agreement No. 605728. MMC acknowledges key support from NSCL as part of the transnational MoU agreement between ENSAR2 and the FRIB Laboratory. NS acknowledges financial support from CNRS/IN2P3 in the framework of the ``Isospin-symmetry breaking:Theory'' Master project.

% references: 
% {Lenzi2001} S.M. Lenzi et al., Phys.Rev.Lett. 87, 122501 (2001)
% {Zuker2002} A.P.Zuker, S.M.Lenzi, G.Martinez-Pinedo, A.Poves, Phys.Rev.Lett. 89, 142502 (2002)
% {Bentley2007} M.A. Bentley and S.M. Lenzi,  Prog.Part.Nucl.Phys. 59, 497 (2007)
% {Bentley2015} M.A.Bentley, S.M.Lenzi, S.A.Simpson, C.Aa.Diget,  Phys.Rev. C 92, 024310 (2015)
% {Boso2018} A. Boso et al,  Phys.Rev.Lett. 121, 032502 (2018)
% {KB3G}A. Poves, J. S\'anchez-Solano, E. Caurier, and F. Nowacki, Nucl. Phys. A 694, 157 (2001).
% {gxpf1a} M. Honma, T. Otsuka, B. A. Brown, and T. Mizusaki, Eur. Phys. J. A 25, 499 (2005).

\bibliographystyle{apsrev4-2}
\bibliography{44V.bib}% Produces the bibliography via BibTeX.

\end{document}